\documentclass[10pt]{article}

\usepackage{amsmath}
\usepackage{amssymb}

\usepackage{graphicx}

\usepackage{cite}

\usepackage{longtable}
\usepackage{multirow}

\usepackage[labelfont=bf,labelsep=period,justification=raggedright]{caption}

\makeatletter
\renewcommand{\@biblabel}[1]{\quad#1.}
\makeatother

\date{}

\begin{document}

% Title must be 150 characters or less
\begin{flushleft}
{\Large
\textbf{A Computational Model of Liver Iron Metabolism}
}
% Insert Author names, affiliations and corresponding author email.
\\
Simon Mitchell$^{1}$, 
Pedro Mendes$^{1,2}$$\ast$ 
\\
\bf{1} School of Computer Science and Manchester Institute of Biotechnology, University of Manchester, Manchester, UK
\\
\bf{2} Virginia Bioinformatics Institute, Virginia Tech, Blacksburg, Virginia, USA
\\

$\ast$ E-mail: pedro.mendes@manchester.ac.uk
\end{flushleft}

% Please keep the abstract between 250 and 300 words
\section*{Abstract}

Iron is essential for all known life due to its redox properties, however these same properties can also lead to its toxicity in overload through the production of reactive oxygen species. Robust systemic and cellular control are required to maintain safe levels of iron and the liver seems to be where this regulation is mainly located. Iron misregulation is implicated in many diseases and as our understanding of iron metabolism improves the list of iron-related disorders grows. Recent developments have resulted in greater knowledge of the fate of iron in the body and have led to a detailed map of its metabolism, however a quantitative understanding at the systems level of how its components interact to produce tight regulation remains elusive.

A mechanistic computational model of human liver iron metabolism, which includes the core regulatory components, is presented here. It was constructed based on known mechanisms of regulation and on their kinetic properties, obtained from several publications. The model was then quantitatively validated by comparing its results with previously published physiological data, and it is able to reproduce multiple experimental findings. A time course simulation following an oral dose of iron was compared to a clinical time course study and the simulation was found to recreate the dynamics and time scale of the systems response to iron challenge. A disease state simulation of haemochromatosis was created by altering a single reaction parameter that mimics a human haemochromatosis gene (HFE) mutation. The simulation provides a quantitative understanding of the liver iron overload that arises in this disease. 

This model supports and supplements understanding of the role of the liver as an iron sensor and provides a framework for further modelling, including simulations to identify valuable drug targets and design of experiments to improve further our knowledge of this system.  

% Please keep the Author Summary between 150 and 200 words
% Use first person. PLoS ONE authors please skip this step. 
% Author Summary not valid for PLoS ONE submissions.   
\section*{Author Summary}
Iron is an essential nutrient required for healthy life but, in excess, is the cause of debilitating and even fatal conditions. The most common genetic disorder in humans caused by a mutation, haemochromatosis, results in an iron overload in the liver. Indeed the liver plays a central role in the regulation of iron. Recently, an increasing amount of detail has been discovered about molecules related to iron metabolism, but an understanding of how they work together and regulate iron levels (in healthy people) or fail to do it (in disease) is still missing. We present a mathematical model of the regulation of liver iron metabolism that provides explanations of its dynamics and allows further hypotheses to be formulated and later tested in experiments. Importantly, the model reproduces accurately the healthy liver iron homeostasis and simulates haemochromatosis showing how the causative mutation leads to iron overload. We investigate how best to control iron regulation and identified reactions that can be 
targets of new medicines to treat iron overload. The model provides a virtual laboratory for investigating iron metabolism and improves understanding of the method by which the liver senses and controls iron levels.

\section*{Introduction}

Iron is an essential element from archaea to complex eukaryotes and man \cite{Aisen2001Chemistry}, and is required for many processes including oxygen transport, DNA synthesis and respiration. Iron deficiency is the most common nutritional deficiency affecting a large proportion of all humans \cite{TussingHumphreys2012Review}. The redox activity which provides iron's utility also means poorly regulated iron metabolism can lead to highly toxic free radicals \cite{Kell2009Iron}. Maintaining the delicate balance of iron requires robust cellular and systemic regulation since both iron deficiency and overload can cause cell death \cite{Hentze2004Balancing}. Recent research has lead to a much greater understanding of the mechanisms controlling iron metabolism and a global view of the interactions between iron-related components is beginning to emerge \cite{Dunn2007Iron, Hower2009General}.

The liver has been proposed to play a central role in the regulation of iron homeostasis \cite{Frazer2003Orchestration} through the action of the recently discovered hormone hepcidin \cite{Park2001Hepcidin}. Hepcidin is expressed predominantly in the liver \cite{Pigeon2001New} and distributed in the serum to control systemic iron metabolism. Hepcidin acts on ferroportin to induce its degradation. Ferroportin is the sole iron exporting protein in mammalian cells \cite{VanZandt2008Iron}, therefore  hepcidin expression reduces iron export into the serum from enterocytes, and reduces iron export from the liver.  Intracellular iron metabolism is controlled by the action of iron response proteins (IRPs) \cite{Hentze1996Molecular}. IRPs post-transcriptionally regulate mRNAs encoding proteins involved in iron metabolism. IRPs combined with ferritin and the transferrin receptors (TfR) make up the center of cellular iron regulation. Ferritin is the iron-storage protein forming a hollow shell which counters 
the toxic effects of free iron by storing iron atoms in a chemically less reactive ferrihydrite \cite{Harrison1977Ferritin}. Extracellular iron circulates bound to transferrin (Tf), and is imported into the cell through the action of membrane bound proteins transferrin receptors 1 and 2 (TfR1 and TfR2). Human haemochromatosis protein (HFE) competes with transferrin bound iron for binding to TfR1 and TfR2 \cite{West2001Mutational}.

Systems Biology provides an excellent methodology for elucidating understanding, through computational modelling, of the complex iron metabolic network. A quantitative model of iron metabolism allows for a careful and principled examination of the effect of the various components of the network. Modelling allows one to do ``what-if'' experiments leading to new hypotheses that can later be put to test experimentally. However, no comprehensive model of liver iron metabolism exists to date. Models have been published that cover specific molecular events only, such as the loading of iron in ferritin \cite{Salgado2010Mathematical}. A qualitative map of mammalian iron metabolism provides a detailed overview of the molecular interactions involved in iron metabolism, including in specific cell types \cite{Hower2009General}. Similarly, a detailed model of iron metabolism and oxidative stress was described but uses a Boolean approach and is specific for yeast \cite{Achcar2011}. Quantitative models of the iron network 
have been recently described
\cite{Chifman2012Core,Mobilia2012}, yet these include only a few components of the iron network. The model from Chifman {\it et al.} suggests that the dynamics of this iron network is stable \cite{Chifman2012Core}. Large-scale models of the metabolism of the hepatocyte \cite{Gille2010,Krauss2012} and a generic human metabolism stoichiometric model \cite{Thiele2013} have also been published, but these contain only four reactions relating to iron metabolism. While they include iron transport, the receptors are not considered, and regulatory details are absent altogether.

Existing models are therefore at two extremes of detail: very specific and very generic --- but to address questions about hepatic iron regulation, what is desirable is a model that balances coverage and detail. This is the aim of the present work. One of the 
problems of modelling iron metabolism quantitatively and in detail arises from the lack of parameter values for many interactions. 
Recently, several of those parameters have been described in the literature (Table \ref{tab:reaction_parameters}), particularly using technologies like surface plasmon resonance. This has enabled us to construct a detailed mechanistic kinetic model of human hepatocyte iron metabolism. The model has been validated by being able to reproduce data from several disease conditions --- importantly, these physiological data were not used in constructing the model. This validation provides a sense of confidence that the model is indeed appropriate for understanding liver iron regulation and for predicting the response to various environmental perturbations.

% Results and Discussion can be combined.
\section*{Results}

\begin{figure}[!ht]
\begin{center}
\includegraphics[width=6.38in]{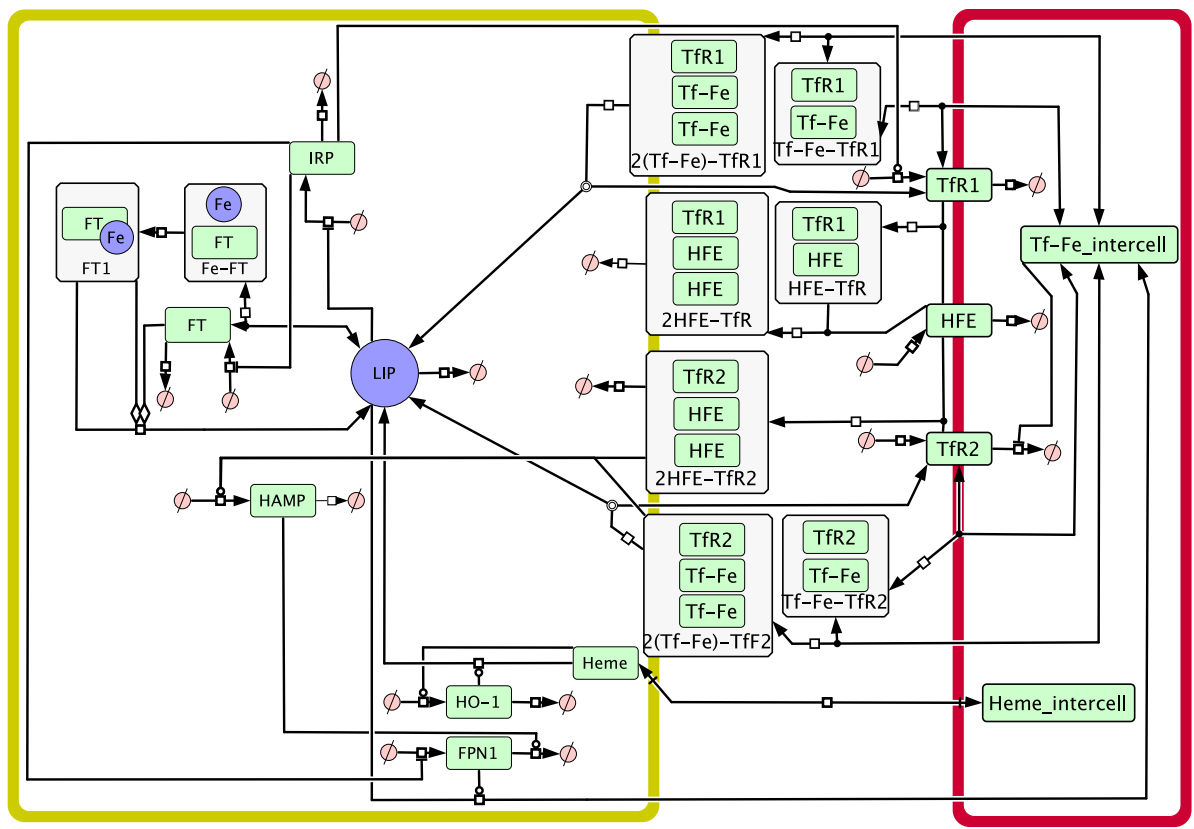}
\end{center}
\caption{
{\bf SBGN process diagram of human liver iron metabolism model.} The compartment with yellow boundary represents the hepatocyte, while the compartment with pink boundary represents plasma. Species overlayed on the compartment boundaries represent membrane-associated species. Abbreviations: Fe: iron, FPN1: ferroportin, FT: ferritin, HAMP: hepcidin,  haeme: intracellular haeme, haeme\_intercell: plasma haeme, HFE: human haemochromatosis protein, HO-1: haeme oxygenase 1, IRP: iron response protein, LIP: labile iron pool,  Tf-Fe\_intercell: plasma transferrin-bound iron, TfR1: transferrin receptor 1, TfR2: transferrin receptor 2. Complexes are represented in boxes with the component species. In the special case of the ferritin-iron complex symbol, the amounts of each species are not in stoichiometric amounts (since there are thousands of iron ions per ferritin).
}
\label{bigDiagram}
\end{figure}

Our model was constructed based on many published data on individual molecular interactions (see Methods section), and is available in Systems Biology Markup Language (SBML) and COPASI formats in supplementary data, as well as from BioModels (http://identifiers.org/biomodels.db/MODEL1302260000) \cite{LeNovere2006BioModels}. Figure \ref{bigDiagram} depicts a process diagram of the model, using the Systems Biology Graphical Notation (SBGN) standard \cite{Novere2009Systems},  where all the considered interactions are shown. It is important to highlight that while results described below are largely in agreement with observations, the model was not forced to replicate them. The extent of agreement between model and physiological data provides confidence that the model is accurate enough to carry out ``what-if'' type of experiments that can provide quantitative explanation of iron regulation in the liver.

\subsection*{Steady State}

Initial validation of the model was performed by assessing the ability to recreate experimentally-observed steady-state concentrations of metabolites and rates of reactions. Simulations were run to steady state using the parameters and initial conditions from Tables \ref{tab:tableOfParams} and \ref{tab:reaction_parameters}. Table \ref{tab:tableOfSS} compares steady-state concentrations of metabolites and reaction rates with experimental observations. 

Chua {\it et al.}  \cite{Chua2010Iron} injected radio-labeled transferrin bound iron into the serum of mice and measured the total uptake of the liver after 120 minutes. The uptake rate, when expressed as mol/s, was close to that found at steady state by the computational model (Table \ref{tab:tableOfSS}).

A technical aspect of note in this steady state solution, is that it is very stiff. This originates because one section of the model is orders or magnitude faster than the rest: the cycle composed of iron binding to ferritin, internalization and release. Arguably this could be resolved by simplifying the model, but the model was left intact because this cycling is an important aspect of iron metabolism and allows the representation of ferritin saturation. Even though the stiffness is high, our software is able to cope by using an appropriate numerical method.

\subsection*{Response to Iron Challenge}
\begin{figure}[!ht]
\begin{center}
\includegraphics[width=3.271in]{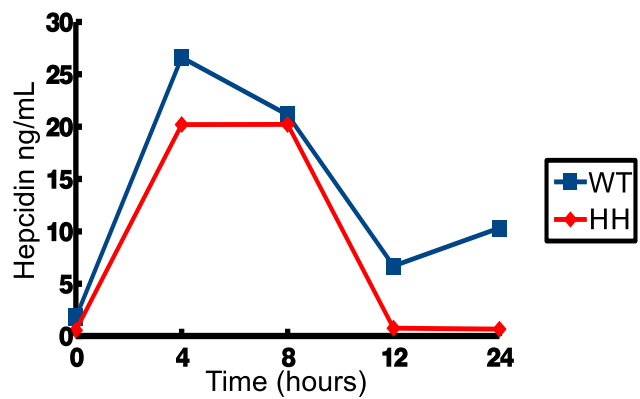}
\end{center}
 \caption{
 {\bf Simulated time course concentrations of hepcidin in response to changing serum transferrin-bound iron levels.} The model shows similar dynamics to time course samples from patients measured by mass spectrometry and ELISA by Girelli {\it et al.}, 2011 \cite{Girelli2011Time}. Hereditary haemochromatosis simulations show reduced hepcidin levels and peak response compared to WT (Wild Type).}

\label{fig:IronResponse}
\end{figure}

An oral dose of iron creates a fluctuation in serum transferrin saturation of approximately 10\% \cite{Girelli2011Time}. The fixed serum iron concentration in the simulation was replaced by a transient increase in concentration equivalent to a 10\% increase in transferrin saturation as a simulation of oral iron dosage on hepatocytes. The simulated hepcidin response (Figure \ref{fig:IronResponse}A) is consistent with the hepcidin response measured by Girelli {\it et al.}  \cite{Girelli2011Time} (Figure \ref{fig:IronResponse}B \& C). The time scale and dynamics of the hepcidin response to iron challenge has been accurately replicated in the simulation presented here. Although the exact dynamics of the simulated response is not validated by either experimental technique (mass spectrometry or ELISA) the simulation appears to present an approximation of the two experimental techniques reaching a peak between 4 and 8 hours and returning to around basal levels within 24 hours.

\subsection*{Cellular Iron Regulation}

\begin{figure}[!ht]
\begin{center}
\includegraphics[width=6.38in]{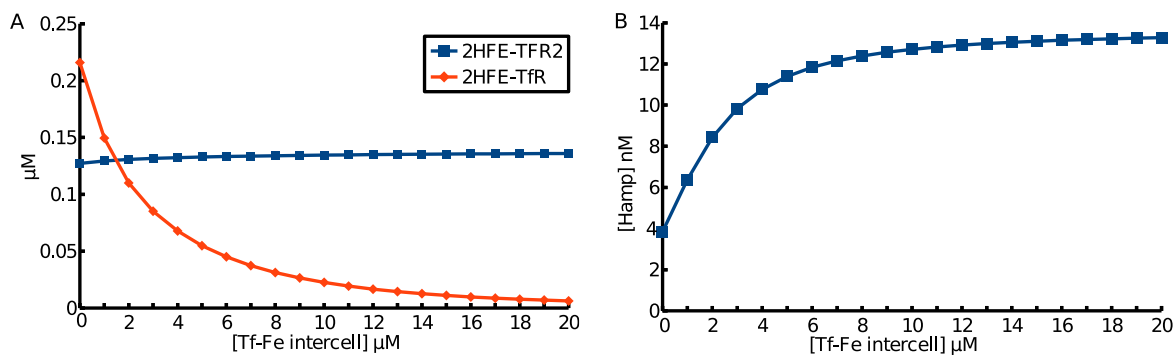}
\end{center}
\caption{
{\bf Simulated steady state concentrations of metabolites in response to increasing serum Tf-Fe.} Increasing HFE-TfR2 complex as a result of HFE-TfR1 reduction induces increased hepcidin.
}
\label{fig:IronChangeHFETfR}
\end{figure}

The computational model supports the proposed role of HFE and TFR2 as sensors of systemic iron. Figure \ref{fig:IronChangeHFETfR}B shows that as the concentration of HFE bound to TfR2 (HFE-TfR2) increases with serum transferrin-bound iron (Tf-Fe\_intercell), at the same time the abundance of HFE bound to TfR1 (HFE-TfR1) decreases. The increase in HFE-TfR2 complex, even though of small magnitude, promotes increased expression of hepcidin (Figure \ref{fig:IronChangeHFETfR}A). It is through this mechanism that liver cells sense serum iron levels and control whole body iron metabolism through the action of hepcidin. Although the labile iron pool increases with serum transferrin-bound iron in this simulation, this is only because the model does not include the action of hepcidin in reducing duodenal export of iron. Expression and secretion of hepcidin will have a global effect of reducing the labile iron pool.

\subsection*{Hereditary Haemochromatosis Simulation}
Hereditary haemochromatosis is the most common hereditary disorder with a prevalence higher than 1 in 500 \cite{Asberg2001Screening}. Type 1 haemochromatosis is the most common and is caused by a mutation in the HFE gene leading to a misregulation of hepcidin and consequent systemic iron overload.

A virtual HFE knockdown was performed by reducing 100-fold the rate constant for HFE synthesis in the model, to create a simulation of type 1 hereditary haemochromatosis. The simulation was run to steady state and results were compared with experimental findings.

\begin{figure}[!ht]
\begin{center}
\includegraphics[width=3.271in]{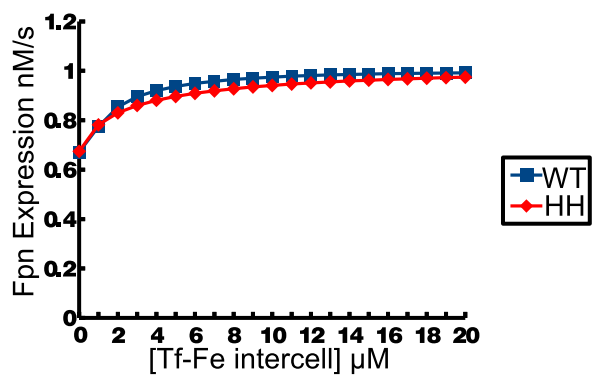}
\end{center}
\caption{
{\bf Ferroportin expression rate in the model doubles in response to changing serum iron concentrations as verified experimentally.} HFE knock-down (HH) simulations and WT simulation of Fe-Tf against ferroportin (Fpn) expression.
}
\label{fig:IronChangeFPN}
\end{figure}

Qualitative validation showed the {\it in silico} HFE knockdown could reproduce multiple experimental findings as shown in Table \ref{table:tableHFEKO}. Quantitatively the model was unable to reproduce accurately the finding that {\it Hfe -/-} mice have 3 times higher hepatic iron levels \cite{Fleming2001Mouse}. This was due to the fixed intercellular transferrin bound iron concentration in the model, unlike in {\it Hfe -/-} mice where there is an increase in transferrin saturation as a result of increased intestinal iron absorption \cite{Fleming2001Mouse}. 
Despite fixed extracellular conditions the model predicted an intracellular hepatocyte iron overload which would be further compounded by the systemic effects of the misregulation of hepcidin. The simulation recreated increased ferroportin levels despite the expression of ferroportin remaining the same as wild type which was consistent with mRNA measurements from Ludwiczek {\it {\it et al.} } \cite{Ludwiczek2005Regulatory}. mRNA based experiments can be used to validate expression rates and protein assays are able to validate steady state protein concentrations as both expression rates and steady state protein concentrations are available as results from the computational model. 
The model of haemochromatosis was also able to replicate the dynamics of experimental responses to changing dietary iron conditions. An approximate 2-fold increase in hepatic ferroportin expression is caused by increased dietary iron in both haemochromatosis and healthy mice \cite{Ludwiczek2005Regulatory}. The model presented here recreated this increase with increasing intercellular iron as can be seen in Figure \ref{fig:IronChangeFPN}.

HFE knockout has been shown to impair the induction of hepcidin by iron in mouse \cite{Ludwiczek2005Regulatory}, and human \cite{Piperno2007Blunted} hepatocytes and this was seen in the computational model as increasing transferrin bound iron did not induce hepcidin as strongly in HFE knockdown.

Although an increase in transferrin receptor 2 was observed in the model ($1.77\mu M$ healthy; $2.80\mu M$ type 1 haemochromatosis), the up-regulation was slightly smaller than the change observed in vivo \cite{Robb2004Regulation}. This is due to the model having fixed extracellular transferrin bound iron concentration, in contrast to haemochromatosis where this concentration increases due to higher absorption in the intestine.

Type 3 haemochromatosis results in similar phenotype as type 1 haemochromatosis, however the mutation is found in the TfR2 gene as opposed to HFE. A virtual TfR2 knockdown mutation was performed by decreasing 100-fold the rate constant of synthesis of TfR2 from the model. Model results were then compared with the findings of Chua {\it et al.}  \cite{Chua2010Iron}. The simulation showed a steady state decrease of liver TfR1 from  $0.29\mu M$ to $0.19\mu M$ with TfR2 knockdown. This is supported by an approximate halving of TfR1 levels in TfR2 mutant mice \cite{Chua2010Iron}. An increase in hepcidin and consequent decrease in ferroportin as seen in mice was matched by the simulation.

An iron overload phenotype with increased intracellular iron is not recreated by the model of the TfR2 mutant. This is, again, due to the fixed serum transferrin-bound iron concentration, while in the whole body there would be increased iron absorption from the diet through the effect of hepcidin.

\subsubsection*{Metabolic Control Analysis}
Metabolic control analysis (MCA) is a standard technique to identify the reactions that have the largest influence on metabolite concentrations or reaction fluxes in a steady state \cite{Kacser1973Control,Heinrich1974Linear}. MCA is a special type of sensitivity analysis and thus is used to quantify the distributed control of the biochemical network. A control coefficient measures the relative change of the variable of interest caused by a small change in the reaction rate (e.g. a control coefficient can be interpreted as the percentage change of the variable given a 1\% change in the reaction rate).

The control over the concentration of the labile iron pool by each of the model reactions can be seen in Table \ref{table:MCA}. The synthesis and degradation of TfR2, TfR1, HFE and the formation of their complexes were found to have the highest control over the labile iron pool. Synthesis and degradation of IRP was also found to have some degree of control, but synthesis and degradation of hepcidin have surprisingly a very small effect on the labile iron pool.

The control over the hepcidin concentration was also measured (Table \ref{table:MCAHepc}), as the ability to control hepatic hepcidin levels could provide therapeutic opportunities to control whole system iron metabolism, due to its action on other tissues. Interestingly, in addition to the expression and degradation of hepcidin itself, the expression of HFE and degradation of HFETfR2 complex have almost as much control over hepcidin. The expression of TfR2 has a considerably lower effect, though still significant.

Flux control coefficients were also determined which indicate the control that reactions have on a chosen reaction flux. The flux control coefficients for the ferroportin mediated iron export reaction are given in Table \ref{table:MCAFlux}. This reaction is of particular interest as it is the only method of iron export, therefore controlling this reaction rate could be important in treating various iron disorders including haemochromatosis and anemia. The reactions of synthesis and degradation of TfR1, TfR2 and HFE were found to have high control, despite not having direct interactions with ferroportin. TfR1 and TfR2 may show consistently high control due to having dual roles as iron importers and iron sensors which control hepcidin expression.

A drawback of MCA, and any other local sensitivity analysis, is that it is only predictive for small changes of reaction rates. However, the changes that result in disease states are usually large, and experimental parameter estimation can result in large uncertainty. Thus a global sensitivity analysis was also performed following the method described in \cite{Sahle2008New}. This calculates the maximal and minimal values of the sensitivity coefficients within a large space of parameter values. This technique is useful, for example, if there is uncertainty about the values of the model parameters as it reveals the possible range of control of each one given the uncertainty. All parameters were allowed to vary within $\pm$ 10\% and the maximal and minimal control coefficients were  measured (Tables \ref{table:MCA}, \ref{table:MCAHepc} and \ref{table:MCAFlux}).

In terms of the control of the labile iron pool (Table \ref{table:MCA}), the reactions with highest control in the reference steady state are still the ones with highest control in the global case ({\it i.e.} when all parameters have an uncertainty of $\pm 10\%$). However TfR1 expression, TfR1 binding, TfR1 degradation, IRP expression and IRP degradation, which all have significant (but not the highest) control in the reference state, could have very low control in the global sense. 
On the other hand HFETfR2 degradation, hepcidin expression, hepcidin degradation and TfR2 binding 2, have low control in the reference steady state, but could have significant control in the global sense. All other reactions have low control in any situation. 

In the case of the control of hepcidin concentration (Table \ref{table:MCAHepc}) the differences between the reference state and the global are much smaller overall, and one could only identify a few reactions that have moderate control in the reference, but could have a bit less in the global sense (TfR2 expression, TfR2 binding, and TfR2 iron internalisation).

In the case of the control of the flux of iron export (Table \ref{table:MCAFlux}), we find some reactions with high control in the reference that could have low control in the global sense: TfR1 expression, TfR1 biding, TfR1 degradation, IRP expression and IRP degradation. Hepcidin expression, hepcidin degradation, and HFETfR2 degradation have almost no control in the reference, but in the global sense they could exert considerable control. This is very similar to the situation of the control of the labile iron pool.

Chifman {\it et al.}  \cite{Chifman2012Core} analysed the parameter space of their core model of iron metabolism in breast epithelial cells and concluded the system behavior is far more dependent on the network structure than the exact parameters used. The analysis presented here lends some support to that finding, since only a few reactions could have different effect on the system if the parameters are wrong. 
A further scan of initial conditions for metabolites found that varying initial concentrations over 2 orders of magnitude had no affect on the steady state achieved (Table \ref{tab:tableOfSS}), indicating that the steady state found in these simulations is unique.

\subsection*{Receptor Properties}
\begin{figure}[!ht]
\begin{center}
\includegraphics[width=3.271in]{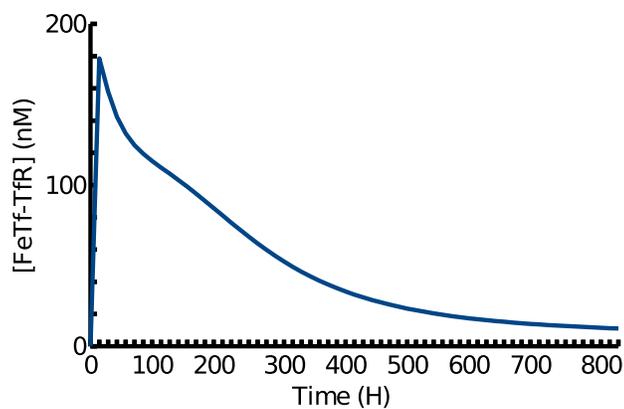}
\end{center}
\caption{
{\bf Iron and Epo receptors show a similar response following an impulse of ligand.} Ligand receptor binding for iron shows a distinctive curve which closely resembles EPO receptor binding studied by Becker {\it et al.} 2010 \cite{Becker2010Covering} (their Fig. 2B).}
\label{fig:EpoIron}
\end{figure}

\begin{figure}[!ht]
\begin{center}
\includegraphics[width=3.271in]{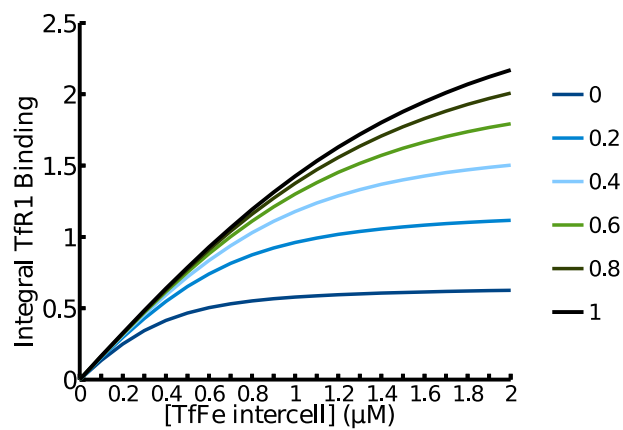}
\end{center}
\caption{
{\bf Increasing receptor turnover increases the linearity of the response for transferrin receptor 1.} The range of linear response for the transferrin receptor depends on its half-life. This effect was first demonstrated in the EPO receptor by Becker {\it et al.} 2010 \cite{Becker2010Covering} who found similar behavior (their Fig. 4D).}
\label{fig:TfRResponse}
\end{figure}

It is known that the iron sensing by the transferrin receptors is responsive over a wide range of intercellular iron concentrations \cite{Lin2007Iron}. The present model reproduces this well (Figure \ref{fig:TfRResponse}A, $1\times$ turnover line). Becker {\it et al.} argued that a linear response of a receptor to its signal over a wide range could be achieved through a combination of: high receptor abundance, increased expression when required, recycling to the surface of internalised receptors and high receptor turnover \cite{Becker2010Covering}. This was illustrated with the behaviour of the erythropoietin (EPO) receptor \cite{Becker2010Covering}. Since the present model contains essentially the same type of reactions that can lead to such a behaviour, simulations were carried out to investigate to what extent this linearity of response is present here. In this case it is the response of the total amount of all forms of TfR1 and TfR2 bound to Tf-Fe against the amount of Tf-Fe\_intercell that 
is important. A variable was created in the model to reflect the total receptor response (see Material and Methods), and this variable was followed in a time course response to an iron pulse (Figure \ref{fig:EpoIron}A). The response to the iron pulse is remarkably similar to the response of the EPO receptor to EPO \cite{Becker2010Covering}. 

\begin{figure}[!ht]
\begin{center}
\includegraphics[width=3.27in]{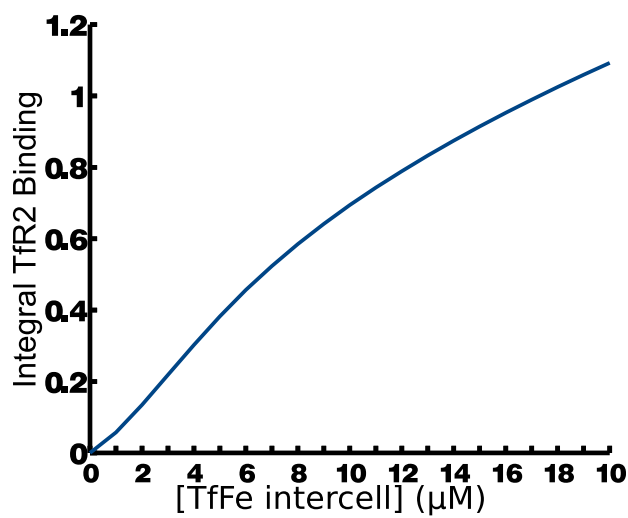}
\end{center}
\caption{
{\bf TfR2 response {\it versus} intercellular transferrin-bound iron.} The response is approximately  linear over a wide range of intercellular iron concentrations.
}
\label{fig:TfR2Response}
\end{figure}

Becker et al. \cite{Becker2010Covering} reported that the linearity of EPO-R response, {\it i.e.} the integral of the response curve, is increased by increasing turnover rate of the receptor and this property was also observed in the simulation of TfR1 response (Figure \ref{fig:TfRResponse}). The range in which the iron response is linear is smaller than that found for EPO (Figure \ref{fig:TfRResponse}). As TfR1's half life in the model matches the experimentally determined value \cite{Chloupkova2010Stoichiometries} the non-linear receptor response seen in the simulation is expected to be accurate. This suggests that TfR1 is a poor sensor for high levels of intercellular iron. On the other hand TfR2 is more abundant than TfR1 \cite{Chloupkova2010Stoichiometries} and accordingly shows an increased linearity for a greater range of intercellular iron concentrations (Figure \ref{fig:TfR2Response}). This suggests the two transferrin receptors play different roles in sensing intercellular iron levels with TfR2 
providing a wide range of 
sensing and TfR1 sensing smaller perturbations. The 
activation of TfR2 directly influences the expression of hepcidin and therefore it is desirable for it to sense large systemic imbalances. TfR1 does not modulate hepcidin expression itself instead it plays a primary role as an iron transporter .

\section*{Discussion}

Iron is an essential element of life, in humans it is involved in oxygen transport, respiration, biosynthesis, detoxification, and other processes. Iron regulation is essential because iron deficiency results in debilitating anemia, while iron excess leads to free radical generation and is involved in many diseases \cite{Kell2009Iron}. It is clear that healthy life depends on tight regulation of iron in the body. The mechanisms involved in iron absortion, transport, storage and regulation form a complex biochemical network \cite{Hower2009General}. The liver has a central role in the regulation of systemic iron metabolism through secretion of the peptide hormone hepcidin. 

Here we analysed the hepatic biochemical network involved in iron sensing and regulation through a mathematical model and computer simulation. The model was constructed mostly based on {\it in vitro} biochemical data, such as protein complex dissociation constants. The model was then validated by comparison with experimental data from multiple physiological studies at both steady state and during dynamic responses. Where quantitative data were available the model matched these well and also qualitatively recreated many findings from clinical and experimental investigations. The simulation accurately modelled the highly prevalent iron disorder haemochromatosis. The disease state was simulated through altering a single parameter of the model and showed quantitatively how an iron overload phenotype occurs in patients with a HFE mutation.

%%new section about translational approach%%
Due to the limited availability of quantitative clinical data on human iron metabolism, various other data sources, particularly from {\it in vitro} experiments and animal models, were integrated for the parameterisation of this model.
This computational modelling effort constitutes a clinical translational approach, enabling data from multiple sources to improve our understanding of human iron metabolism. Several arguments could be raised to cast doubt on this approach, such as the the failure of {\it in vitro} conditions to mimic those {\it in vivo} or the difference between animal models and humans. This means that this type of data integration must be carefully monitored in terms of establishing the validity of the resulting model. Examining the behaviour of the model by simulating it at different values of initial conditions or other parameters (parameter scans) is important to establish the limits of utility of the model. Global sensitivity analysis is another approach that 
determines the boundaries of parameter variation that the model tolerates before it becomes too distant from the actual system behaviour. Model validation is an essential step in modelling. Validation should be carried out by enquiring if the model is able to match experimental observations that were not used to calibrate it. Here the model was validated by the simulation of haemochromatosis disease, where the model behaviour matched the clinical data (Table \ref{table:tableHFEKO}).

The precise regulatory mechanism behind transferrin receptors and HFE controlling hepcidin expression remains to be validated experimentally, however the model presented here supports current understanding that the interaction of TfR2 and HFE form the signal transduction pathway that leads to the induction of hepcidin expression \cite{Gao2009Interaction}.

The global metabolic control analysis results support the identification of the transferrin receptors, particularly TfR2, and HFE as potential therapeutic targets; a result that is robust even to inaccuracies in parameter values. Although hepcidin would be an intuitive point of high control of this system (and therefore a good therapeutic target), in the present model this is not the case. It seems that targeting the promoters of hepcidin expression may be more desirable. However this conclusion has to be expressed with some reservation that stems from  the fact that the global sensitivity analysis identified the hepcidin synthesis and degradation reactions in the group of those with the largest uncertainty. By changing parameter values by no more than 10\% it would be possible to have the hepcidin expression and degradation show higher control. So it seems important that the expression of hepcidin be studied in more detail. We also predict that the control of hepcidin over the system would be 
higher if the 
model had included the regulation of intestinal ferroportin by hepatic ferroportin.

The global sensitivity analysis, however, allows taking strong conclusions about the reactions for which the reference steady state is not much different from the maximal and minimal values. It turns out that these are the reactions that have the largest and the smallest control over the system variables. For example, the reactions with greatest control on the labile iron pool and iron export are those of the HFE-TfR2 system. But the reactions of the HFE-TfR1 system have always low control. These two conclusions are valid {\it under a wide range of parameter values}.

Construction of this model required several assumptions to be made due to lack of measured parameter values, as described in Material and Methods. These assumptions may or may not have a large impact on the model behaviour, and it is important to identify those that have a large impact, as their measurement will improve our knowledge the most. Of all the assumptions made, the rates of expression and degradation of ferroportin are those that have a significant impact on the labile iron pool in the model (see Table \ref{table:MCA}). This means that if the values assumed for these rate parameters were to be significantly different the model prediction for labile iron pool behaviour would also be different. The model is therefore also useful by suggesting experiments that will optimally improve our knowledge about this system.

Limitations on the predictive power of the model occur due to the scope of the system chosen. Fixed serum iron conditions, which were used as boundary conditions in the model, do not successfully recreate the amplifying feedbacks that occur as a result of hepcidin expression controlling enterocyte iron export. To relieve this limitation, a more advanced model should include dietary iron uptake and the action of hepcidin on that process.
 
The model predicts a quasi-linear response to increasing pulses of serum iron, similar to what has been predicted for the erythropoietin system \cite{Becker2010Covering}.  Our simulations   display response of the transferrin receptors to pulses of extracellular transferrin-bound iron that is similar to the EPO-R response to EPO (Figure \ref{fig:TfRResponse}). The integral of this response {\it versus} the iron sensed deviates very little from linearity in the range of physiological iron (Figure \ref{fig:EpoIron}). 

Computational models are research tools whose function is to allow for reasoning in a complex nonlinear system. The present model can be useful in terms of predicting properties of the liver iron system. These predictions form hypotheses that lead to new experiments. Their outcome will undoubtedly improve our knowledge and will also either confirm the accuracy of the model or refute it (in which case it then needs to be corrected). The present model and its results identified a number of predictions about liver iron regulation that should be investigated further:
\begin{itemize}
\item changes in activity of the hepcidin gene in the liver have little effect on the size of the labile iron pool,
\item the rate of expression of HFE has a high control over the steady state level of hepcidin,
\item the strong effect of HFE is due to its interaction with TfR2 rather than TfR1,
\item the rate of liver iron export by ferroportin has a strong dependence on the expression of TfR1, TfR2 and HFE,
\item the rate of expression of hepcidin is approximately linear with the concentration of plasma iron within the physiological range.
\end{itemize}

The present model is the most detailed quantitative mechanistic model of cellular iron metabolism to date, allowing for a comprehensive description of its regulation. It can be used to elucidate the link from genotype to phenotype, as demonstrated here with hereditary haemochromatosis. The model provides the ability to investigate scenarios for which there are currently no experimental data available --- thus making predictions about the system and aiding in experimental design.

% You may title this section "Methods" or "Models". 
% "Models" is not a valid title for PLoS ONE authors. However, PLoS ONE
% authors may use "Analysis" 
\section*{Materials and Methods}

The model is constructed using ordinary differential equations (ODEs) to represent the rate of change of each chemical species. COPASI \cite{Hoops2006COPASIa} was used as the software framework for model construction, simulation and analysis. Cell Designer \cite{Funahashi2003CellDesigner} was used for construction of a SBGN process diagram (Figure \ref{bigDiagram}).

The model consists of 2 compartments representing the  serum and the liver. 
Concentrations of haeme and transferrin bound iron in the serum were fixed to represent constant extracellular conditions. Fixed metabolites simulate a constant influx of iron through the diet as any iron absorbed by the liver is effectively replenished. A labile iron pool (LIP) consumption reaction is added to represent various uses of iron and create a flow through the system. 
%% new sentence about exclusion of heme biosynthesis %%
Some of the LIP consumption reaction would be attributed to heme biosynthesis however this process was not considered explicitly in this study to avoid unnecessary complexity and because the bone marrow is the major site of heme biosynthesis \cite{Ajioka2016}.
%% end of new section about exclusion of heme biosynthesis %%

Initial concentrations for metabolites were set to appropriate concentrations based on a literature survey (Table \ref{tab:tableOfParams}). All metabolites formed through complex binding were set to zero initial concentrations (Table \ref{tab:tableOfParams}).
The concentration of a chemical species at a time point in the simulation is determined by integrating the system of ODEs. For some proteins a half-life was available in the literature, but sources could not be found for their synthesis rates (translation). In this occurrence, estimated steady-state concentrations were used from the literature and a synthesis rate was chosen such that at steady state the concentration of the protein would be approximately accurate, following Equation \ref{eq:synthrate}:

\begin{equation} 
 \label{eq:synthrate}
\frac{d[\textrm{P]}}{dt} = +k-d[\textrm{P}] = 0.
 \end{equation}

This is solved for $k$ where $[\textrm{P}]$ is the steady-state concentration of the protein and $d$ is the degradation rate obtained from the half-life ($\lambda$) using:

\begin{equation} 
 \label{eq:lambda}
d=\frac{\ln{2}}{\lambda}.
 \end{equation}

Complex formation reactions, such as binding of TfR1 to Tf-Fe for iron uptake, are modelled using the on and off binding constants as a forward and reverse mass action reaction. For example:

\begin{equation} 
 \label{eq:rev}
\textrm{TfR1}  +  \textrm{Tf-Fe} \rightleftharpoons \textrm{Tf-Fe-TfR1}
\end{equation}

is modelled using two reactions:
\begin{equation} 
 \label{eq:irrevf}
\textrm{TfR1}  +  \textrm{Tf-Fe} \stackrel{k_a}{\rightarrow} \textrm{Tf-Fe-TfR1}
\end{equation}

\begin{equation} 
 \label{eq:irrevr}
\textrm{Tf-Fe-TfR1} \stackrel{k_d}{\rightarrow} \textrm{TfR1}  +  \textrm{Tf-Fe} 
\end{equation}

There is one ODE per each chemical species. The two reactions \ref{eq:irrevf} and \ref{eq:irrevr} add the following terms to the set of ODEs:

\begin{equation} 
 \label{eq:ode}
\begin{split}
\frac{d[\textrm{TfR1}]}{dt} =&  -k_a[\textrm{TfR1}][\textrm{TF-Fe}]+k_d[\textrm{Tf-Fe-TfR1}] \ldots\\
\frac{d[\textrm{Tf-Fe}]}{dt} =& -k_a[\textrm{TfR1}][\textrm{TF-Fe}]+k_d[\textrm{Tf-Fe-TfR1}] \ldots\\
\frac{d[\textrm{Tf-Fe-TfR1}]}{dt} =& +k_a[\textrm{TfR1}][\textrm{TF-Fe}]-k_d[\textrm{Tf-Fe-TfR1}] \ldots
\end{split}
\end{equation}

Intracellular haeme levels are controlled by a balance between uptake, export and oxygenation. haeme import through the action of haeme carrier protein 1 (HCP1), export by ATP-binding cassette sub-family G member 2 (ABCG2) and oxygenation by haeme oxygenase-1 (HO-1) follow Michaelis-Menten kinetics. HO-1 expression is promoted by haeme through by a Hill function (Equation \eqref{eqn:Hill}).
\begin{align}
v = [\textrm{S}] \cdot a \cdot &\left( \frac{[\textrm{M}]^{n_{H}}}  {K^{n_{H}}+[\textrm{M}]^{n_{H}}} \right),\label{eqn:Hill}\\
v = [\textrm{S}] \cdot a \cdot &\left( 1-\frac{[\textrm{M}]^{n_{H}}}{K^{n_{H}}+[\textrm{M}]^{n_{H}}} \right).\label{eqn:invHill}
\end{align}
Where $S$ is the substrate, $M$ is the modifier, $a$ is the turnover number, $K$ is the ligand concentration which produces half occupancy of the binding sites of the enzyme, and $n_{H}$ is the Hill coefficient. Values of $n_{H}$ larger than $1$ produce positive cooperativity ({\it i.e.} a sigmoidal response); when $n_{H}=1$ the response is the same as Michaelis-Menten kinetics.  A Hill coefficient of $n_{H}=1$ was assumed unless there is literature evidence for a different value. Where $K$ is not known it has been estimated to be of the order of magnitude of experimentally observed concentrations for the ligand.

IRP/Iron-responsive elements (IRE) regulation is represented by Hill kinetics using Equation \eqref{eqn:Hill} to simulate the 3' binding of IRP promoting the translation rate, and Equation \eqref{eqn:invHill} to represent the 5' binding of IRP reducing the translation rate.
Ferroportin degradation is modelled using 2 reactions: one representing the standard half-life and the other representing the hepcidin-induced degradation. A Hill equation (Equation \ref{eqn:Hill}) is used to simulate the hepcidin-induced degradation of ferroportin. 

Hepcidin expression is the only reaction modelled using a Hill coefficient greater than 1. Due to the small dynamic range of HFE-TfR2 concentrations a Hill coefficient of 5 was chosen to provide the sensitivity required to produce the expected range of hepcidin concentrations. The mechanism by which HFE-TfR2 interactions induce hepcidin expression is not well understood, but is thought to involve the mitogen-activated protein kinase (MAPK) signalling pathway \cite{Wallace2009Combined}. The stimulus/response curve of the MAPK cascade has been found to be equivalent to a cooperative enzyme with a Hill coefficient of 4-5 \cite{Huang1996Ultrasensitivity}, making the steep Hill function appropriate to model hepcidin expression.

Ferritin modelling follows the work of Salgado {\it et al.} \cite{Salgado2010Mathematical}. Iron from the LIP binds to, and is internalised in, ferritin with mass action kinetics. Internalised iron release from ferritin occurs through 2 reactions (intact ferritin release and release due to ferritin degradation). The average amount of iron internalised per ferritin affects the iron release rate and this is modelled using Equation \ref{eq:salgado} (adapted from \cite{Salgado2010Mathematical}):

\begin{equation}
 \label{eq:salgado}
v = [\textrm{S}] \cdot k_\textrm{loss} \cdot \left(1+\frac{0.048 \cdot \frac{[\textrm{FT1}]}{[\textrm{FT}]}}{1+\frac{[\textrm{FT1}]}{[\textrm{FT}]}}\right).
\end{equation}

Where $S$ is internalised iron, $k_\textrm{loss}$ is the rate constant and $[\textrm{FT1}]/ [\textrm{FT}]$ is the ratio of iron internalised in ferritin to total ferritin available. The value 0.0048 was obtained by dividing the value given in Salgado {\it et al.} \cite{Salgado2010Mathematical} by 50 as that simulation was scaled for 50 iron atom packages.

Iron is also released from ferritin when the entire ferritin cage is degraded. The kinetics of ferritin degradation are mass action, however the amount of iron released when a ferritin cage is degraded is an average based on ferritin levels and total iron internalised in ferritin. Incorporating mass action and ferritin saturation ratio gives the following rate law for $\textrm{FT1}~\rightarrow~\textrm{LIP}$:

\begin{equation}
v = [\textrm{S}] \cdot k \cdot \frac{[\textrm{FT1}]}{[\textrm{FT}]}.
\end{equation}

Iron export rate was modelled using Equation \ref{eqn:Hill} with ferroportin as the modifier and a Hill coefficient of 1. $K$ was assumed to be around the steady state concentration of IRP ($1\mu M$). A rate (V) of $40\textrm{pm} \cdot (10^6 cells \cdot 5\textrm{min})^{-1}$ was used from Sarkar {\it et al.} \cite{Sarkar2003} and these values were substituted into the equation and solved for $a$.

%%%new section from reviews about parameter estimation%%
Ferroportin expression rates and degradation rates are poorly understood. Ferroportin abundance data \cite{Wang2012PaxDb} lead to an estimate of ferroportin concentration around $0.16 \mu M$. The hepcidin induced degradation of ferroportin is represented in the model by a rate law in the form of Equation \ref{eqn:Hill} with a Hill coefficient $n_{H} = 5$ (see above) and a $K^{n_{H}}$ equal to the measured concentration of hepcidin \cite{Zaritsky2010Reduction} (see Table \ref{tab:tableOfParams}). We then assume a maximal rate of degradation to be 1 $nM s^{-1}$, and using the steady state concentration of ferroportin, the rate constant can be estimated as 0.0002315 $s^{-1}$. The ferroportin synthesis rate was then calculated to produce the required steady-state concentration of ferroportin at the nominal hepcidin concentration.

The HFE-TfR2 binding and dissociation constants were also not available and so it was assumed that they were the same as those of TfR1-HFE. Finally, the HFE-TfR and HFE-TfR2 degradation rates are also not known; we used a value that is an order of magnitude lower than the half life for unbound TfR (i.e. we assume the complex to be more stable than the free form of TfR).
%%end of new section from reviews%%

%%new section explaining exclusion of DMT1%%
Although DMT1 may contribute towards transferrin bound iron uptake in hepatocytes this contribution has been found to be minor and DMT1 knockout has little affect on iron metabolism \cite{Wang2013Hepatocyte}, therefore DMT1 was not included in the model.
%%end of new section explaining exclusion of DMT1%%

%%new section explaining IRPs%%
The two iron response proteins (IRP1 and IRP2) which are responsible for cellular iron regulation were modelled as a single metabolite in this study as the mechanistic differences in their regulatory roles is poorly understood. Equivalent regulation by both IRPs has been found in multiple studies \cite{Kim1995Translational, Ke1998Loops, Erlitzki2002Multiple}.
%%end of new section explaining IRPs%%

Global sensitivity analysis was performed using the method proposed by Sahle {\it et al.} \cite{Sahle2008New}, where all parameter values were allowed to vary within $\pm 10\%$ of their nominal value in the model and we search for the maximum and minimum value that concentration- or flux-control coefficients of interest are able to reach within that parameter space. The searches were carried out with the particle swarm optimisation algorithm \cite{494215}. In order to process these optimisations in a reasonable time  a HTCondor \cite{condor-hunter} distributed computing system was used, managed through the Condor-COPASI package \cite{22834945}.

To perform analysis of receptor response in a similar manner to the EPO system studied by Becker {\it et al.} \cite{Becker2010Covering} initial conditions were adjusted to recreate a similar virtual experiment. Haeme concentration was fixed at zero to isolate transferrin bound iron uptake. The rate constant of the labile iron pool depletion reaction was reduced to balance the reduced iron uptake (which results in iron having a similar half-life to EPO in \cite{Becker2010Covering}). Initial concentrations for all metabolites were set to steady-state concentrations with the exception of the labile iron pool and iron bound to all receptors which were set to zero. Extracellular transferrin bound iron was allowed to vary and set at increasing concentrations to determine receptor response. Time courses were calculated for Tf-Fe-TfR1, 2(Tf-Fe)-TfR1, Tf-Fe-TfR2 and 2(Tf-Fe)-TfR2 as iron binds its two receptors in a two-staged process. Two new variables were defined in COPASI which integrated the results of the time 
courses corresponding to the two receptors (in their different ligand states): 

\begin{equation}
\textrm{Int\_TfR1\_binding} = \int [\textrm{Tf\-Fe\-TfR1}] \cdot dt + \int [\textrm{2(Tf\-Fe)\-TfR1}] \cdot dt ,
\end{equation}

\begin{equation}
\textrm{Int\_TfR2\_binding} = \int [\textrm{Tf\-Fe\-TfR2}] \cdot dt + \int [\textrm{2(Tf\-Fe)\-TfR2}] \cdot dt .
\end{equation}

% Do NOT remove this, even if you are not including acknowledgments
\section*{Acknowledgments}

We thank Steve Ackman, Douglas Kell, Reinhard Laubenbacher, Frank Torti and Suzy Torti for many discussions. We thank Ed Kent and the EPS IT Research Services staff for help with running the HTCondor pool. We also thank Anthony West for sharing binding data for HFE, TfR1 and TfR2. SM is grateful to the Virginia Bioinformatics Institute and the Wake Forest University Department of Cancer Biology for hosting visits. 

%\section*{References}
% The bibtex filename
\bibliography{refs}

\begin{thebibliography}{10}
\providecommand{\url}[1]{\texttt{#1}}
\providecommand{\urlprefix}{URL }
\expandafter\ifx\csname urlstyle\endcsname\relax
  \providecommand{\doi}[1]{doi:\discretionary{}{}{}#1}\else
  \providecommand{\doi}{doi:\discretionary{}{}{}\begingroup
  \urlstyle{rm}\Url}\fi
\providecommand{\bibAnnoteFile}[1]{%
  \IfFileExists{#1}{\begin{quotation}\noindent\textsc{Key:} #1\\
  \textsc{Annotation:}\ \input{#1}\end{quotation}}{}}
\providecommand{\bibAnnote}[2]{%
  \begin{quotation}\noindent\textsc{Key:} #1\\
  \textsc{Annotation:}\ #2\end{quotation}}
\providecommand{\eprint}[2][]{\url{#2}}

\bibitem{Aisen2001Chemistry}
Aisen P, Enns C, Wessling-Resnick M (2001) {Chemistry and biology of eukaryotic
  iron metabolism.}
\newblock Int J Biochem Cell Biol 33: 940--959.
\bibAnnoteFile{Aisen2001Chemistry}

\bibitem{TussingHumphreys2012Review}
Tussing-Humphreys L, Pustacioglu C, Nemeth E, Braunschweig C (2012) Rethinking
  iron regulation and assessment in iron deficiency, anemia of chronic disease,
  and obesity: introducing hepcidin.
\newblock J Acad Nutrition Dietetics 112: 391--400.
\bibAnnoteFile{TussingHumphreys2012Review}

\bibitem{Kell2009Iron}
Kell D (2009) Iron behaving badly: inappropriate iron chelation as a major
  contributor to the aetiology of vascular and other progressive inflammatory
  and degenerative diseases.
\newblock BMC Med Genomics 2: 2.
\bibAnnoteFile{Kell2009Iron}

\bibitem{Hentze2004Balancing}
Hentze MW, Muckenthaler MU, Andrews NC (2004) {Balancing acts: molecular
  control of mammalian iron metabolism.}
\newblock Cell 117: 285--297.
\bibAnnoteFile{Hentze2004Balancing}

\bibitem{Dunn2007Iron}
Dunn LL, Rahmanto YSS, Richardson DR (2007) {Iron uptake and metabolism in the
  new millennium.}
\newblock Trends Cell Biol 17: 93--100.
\bibAnnoteFile{Dunn2007Iron}

\bibitem{Hower2009General}
Hower V, Mendes P, Torti FM, Laubenbacher R, Akman S, et~al. (2009) {A general
  map of iron metabolism and tissue-specific subnetworks.}
\newblock Mol BioSystems 5: 422--443.
\bibAnnoteFile{Hower2009General}

\bibitem{Frazer2003Orchestration}
Frazer DM, Anderson GJ (2003) {The orchestration of body iron intake: how and
  where do enterocytes receive their cues?}
\newblock Blood Cells, Molecules \& Diseases 30: 288--297.
\bibAnnoteFile{Frazer2003Orchestration}

\bibitem{Park2001Hepcidin}
Park CH, Valore EV, Waring AJ, Ganz T (2001) {Hepcidin, a urinary antimicrobial
  peptide synthesized in the liver.}
\newblock J Biol Chem 276: 7806--7810.
\bibAnnoteFile{Park2001Hepcidin}

\bibitem{Pigeon2001New}
Pigeon C, Ilyin G, Courselaud B, Leroyer P, Turlin B, et~al. (2001) {A new
  mouse liver-specific gene, encoding a protein homologous to human
  antimicrobial peptide hepcidin, is overexpressed during iron overload.}
\newblock J Biol Chem 276: 7811--7819.
\bibAnnoteFile{Pigeon2001New}

\bibitem{VanZandt2008Iron}
Van~Zandt KE, Sow FB, Florence WC, Zwilling BS, Satoskar AR, et~al. (2008) The
  iron export protein ferroportin 1 is differentially expressed in mouse
  macrophage populations and is present in the mycobacterial-containing
  phagosome.
\newblock J Leukocyte Biol 84: 689--700.
\bibAnnoteFile{VanZandt2008Iron}

\bibitem{Hentze1996Molecular}
Hentze MW, K\"{u}hn LC (1996) {Molecular control of vertebrate iron metabolism:
  mRNA-based regulatory circuits operated by iron, nitric oxide, and oxidative
  stress.}
\newblock Proc Natl Acad Sci USA 93: 8175--8182.
\bibAnnoteFile{Hentze1996Molecular}

\bibitem{Harrison1977Ferritin}
Harrison PM (1977) {Ferritin: an iron-storage molecule.}
\newblock Sem Hematol 14: 55--70.
\bibAnnoteFile{Harrison1977Ferritin}

\bibitem{West2001Mutational}
West AP, Giannetti AM, Herr AB, Bennett MJ, Nangiana JS, et~al. (2001)
  {Mutational analysis of the transferrin receptor reveals overlapping HFE and
  transferrin binding sites.}
\newblock J Mol Biol 313: 385--397.
\bibAnnoteFile{West2001Mutational}

\bibitem{Salgado2010Mathematical}
Salgado JC, Nappa AO, Gerdtzen Z, Tapia V, Theil E, et~al. (2010) {Mathematical
  modeling of the dynamic storage of iron in ferritin}.
\newblock BMC Systems Biol 4: 147.
\bibAnnoteFile{Salgado2010Mathematical}

\bibitem{Achcar2011}
Achcar F, Camadro JM, Mestivier D (2011) A boolean probabilistic model of
  metabolic adaptation to oxygen in relation to iron homeostasis and oxidative
  stress.
\newblock BMC Systems Biology 5: 51.
\bibAnnoteFile{Achcar2011}

\bibitem{Chifman2012Core}
Chifman J, Kniss A, Neupane P, Williams I, Leung B, et~al. (2012) The core
  control system of intracellular iron homeostasis: a mathematical model.
\newblock J Theoret Biol 300: 91--99.
\bibAnnoteFile{Chifman2012Core}

\bibitem{Mobilia2012}
Mobilia N, Donz\'{e} A, Moulis JM, Fanchon E (2012) A model of the cellular
  iron homeostasis network using {Semi-Formal} methods for parameter space
  exploration.
\newblock Electronic Proceedings in Theoretical Computer Science 92: 42--57.
\bibAnnoteFile{Mobilia2012}

\bibitem{Gille2010}
Gille C, Bolling C, Hoppe A, Bulik S, Hoffmann S, et~al. (2010) {HepatoNet1}: a
  comprehensive metabolic reconstruction of the human hepatocyte for the
  analysis of liver physiology.
\newblock Molecular Systems Biology 6: 411.
\bibAnnoteFile{Gille2010}

\bibitem{Krauss2012}
Krauss M, Schaller S, Borchers S, Findeisen R, Lippert J, et~al. (2012)
  Integrating cellular metabolism into a multiscale {Whole-Body} model.
\newblock PLoS Comput Biol 8: e1002750.
\bibAnnoteFile{Krauss2012}

\bibitem{Thiele2013}
Thiele I, Swainston N, Fleming RMT, Hoppe A, Sahoo S, et~al. (2013) A
  community-driven global reconstruction of human metabolism.
\newblock Nature Biotechnology 31: 419--425.
\bibAnnoteFile{Thiele2013}

\bibitem{LeNovere2006BioModels}
Le~Nov\`{e}re N, Bornstein B, Broicher A, Courtot M, Donizelli M, et~al. (2006)
  {BioModels} database: a free, centralized database of curated, published,
  quantitative kinetic models of biochemical and cellular systems.
\newblock Nucleic Acids Res 34: D689--D691.
\bibAnnoteFile{LeNovere2006BioModels}

\bibitem{Novere2009Systems}
Novere NL, Hucka M, Mi H, Moodie S, Schreiber F, et~al. (2009) {The Systems
  Biology Graphical Notation}.
\newblock Nature Biotechnol 27: 735--741.
\bibAnnoteFile{Novere2009Systems}

\bibitem{Chua2010Iron}
Chua AC, Delima RD, Morgan EH, Herbison CE, Tirnitz-Parker JE, et~al. (2010)
  Iron uptake from plasma transferrin by a transferrin receptor 2 mutant mouse
  model of haemochromatosis.
\newblock J Hepatol 52: 425--431.
\bibAnnoteFile{Chua2010Iron}

\bibitem{Girelli2011Time}
Girelli D, Trombini P, Busti F, Campostrini N, Sandri M, et~al. (2011) A time
  course of hepcidin response to iron challenge in patients with hfe and tfr2
  hemochromatosis.
\newblock Haematologica 96: 500--506.
\bibAnnoteFile{Girelli2011Time}

\bibitem{Asberg2001Screening}
Asberg (2001) Screening for hemochromatosis: High prevalence and low morbidity
  in an unselected population of 65,238 persons.
\newblock Scandinavian J Gastroenterol 36: 1108--1115.
\bibAnnoteFile{Asberg2001Screening}

\bibitem{Fleming2001Mouse}
Fleming RE, Holden CC, Tomatsu S, Waheed A, Brunt EM, et~al. (2001) Mouse
  strain differences determine severity of iron accumulation in hfe knockout
  model of hereditary hemochromatosis.
\newblock Proc Natl Acad Sci USA 98: 2707--2711.
\bibAnnoteFile{Fleming2001Mouse}

\bibitem{Ludwiczek2005Regulatory}
Ludwiczek S, Theurl I, Bahram S, Sch\"{u}mann K, Weiss G (2005) Regulatory
  networks for the control of body iron homeostasis and their dysregulation in
  hfe mediated hemochromatosis.
\newblock J Cell Physiol 204: 489--499.
\bibAnnoteFile{Ludwiczek2005Regulatory}

\bibitem{Piperno2007Blunted}
Piperno A, Girelli D, Nemeth E, Trombini P, Bozzini C, et~al. (2007) Blunted
  hepcidin response to oral iron challenge in hfe-related hemochromatosis.
\newblock Blood 110: 4096--4100.
\bibAnnoteFile{Piperno2007Blunted}

\bibitem{Robb2004Regulation}
Robb A, Wessling-Resnick M (2004) {Regulation of transferrin receptor 2 protein
  levels by transferrin.}
\newblock Blood 104: 4294--4299.
\bibAnnoteFile{Robb2004Regulation}

\bibitem{Kacser1973Control}
Kacser H, Burns JA (1973) The control of flux.
\newblock Symposia of the Society for Experimental Biology 27: 65--104.
\bibAnnoteFile{Kacser1973Control}

\bibitem{Heinrich1974Linear}
Heinrich R, Rapoport TA (1974) A linear steady-state treatment of enzymatic
  chains.
\newblock European Journal of Biochemistry 42: 89--95.
\bibAnnoteFile{Heinrich1974Linear}

\bibitem{Sahle2008New}
Sahle S, Mendes P, Hoops S, Kummer U (2008) A new strategy for assessing
  sensitivities in biochemical models.
\newblock Phil Trans R Soc A 366: 3619--3631.
\bibAnnoteFile{Sahle2008New}

\bibitem{Lin2007Iron}
Lin L, Valore EV, Nemeth E, Goodnough JB, Gabayan V, et~al. (2007) Iron
  transferrin regulates hepcidin synthesis in primary hepatocyte culture
  through hemojuvelin and bmp2/4.
\newblock Blood 110: 2182--2189.
\bibAnnoteFile{Lin2007Iron}

\bibitem{Becker2010Covering}
Becker V, Schilling M, Bachmann J, Baumann U, Raue A, et~al. (2010) Covering a
  broad dynamic range: Information processing at the erythropoietin receptor.
\newblock Science 328: 1404--1408.
\bibAnnoteFile{Becker2010Covering}

\bibitem{Chloupkova2010Stoichiometries}
Chloupkov\'{a} M, Zhang AS, Enns CA (2010) Stoichiometries of transferrin
  receptors 1 and 2 in human liver.
\newblock Blood Cells, Molecules, \& Diseases 44: 28--33.
\bibAnnoteFile{Chloupkova2010Stoichiometries}

\bibitem{Gao2009Interaction}
Gao J, Chen J, Kramer M, Tsukamoto H, Zhang ASS, et~al. (2009) Interaction of
  the hereditary hemochromatosis protein hfe with transferrin receptor 2 is
  required for transferrin-induced hepcidin expression.
\newblock Cell Metabolism 9: 217--227.
\bibAnnoteFile{Gao2009Interaction}

\bibitem{Hoops2006COPASIa}
Hoops S, Sahle S, Gauges R, Lee C, Pahle J, et~al. (2006) {COPASI --- a COmplex
  PAthway SImulator}.
\newblock Bioinformatics 22: 3067--3074.
\bibAnnoteFile{Hoops2006COPASIa}

\bibitem{Funahashi2003CellDesigner}
Funahashi A, Morohashi M, Kitano H, Tanimura N (2003) {CellDesigner: a process
  diagram editor for gene-regulatory and biochemical networks}.
\newblock BIOSILICO 1: 159--162.
\bibAnnoteFile{Funahashi2003CellDesigner}

\bibitem{Ajioka2016}
Ajioka RS, Phillips JD, Kushner JP (2006) {Biosynthesis of heme in mammals.}
\newblock Biochimica et Biophysica Acta 1763: 723--736.
\bibAnnoteFile{Ajioka2016}

\bibitem{Wallace2009Combined}
Wallace DF, Summerville L, Crampton EM, Frazer DM, Anderson GJ, et~al. (2009)
  Combined deletion of hfe and transferrin receptor 2 in mice leads to marked
  dysregulation of hepcidin and iron overload.
\newblock Hepatol 50: 1992--2000.
\bibAnnoteFile{Wallace2009Combined}

\bibitem{Huang1996Ultrasensitivity}
Huang CY, Ferrell JE (1996) Ultrasensitivity in the mitogen-activated protein
  kinase cascade.
\newblock Proc Natl Acad Sci USA 93: 10078--10083.
\bibAnnoteFile{Huang1996Ultrasensitivity}

\bibitem{Sarkar2003}
Sarkar J, Seshadri V, Tripoulas NA, Ketterer ME, Fox PL (2003) Role of
  ceruloplasmin in macrophage iron efflux during hypoxia.
\newblock The Journal of biological chemistry 278: 44018--44024.
\bibAnnoteFile{Sarkar2003}

\bibitem{Wang2012PaxDb}
Wang M, Weiss M, Simonovic M, Haertinger G, Schrimpf SP, et~al. (2012) Paxdb, a
  database of protein abundance averages across all three domains of life.
\newblock Molecular \& cellular proteomics 11: 492--500.
\bibAnnoteFile{Wang2012PaxDb}

\bibitem{Zaritsky2010Reduction}
Zaritsky J, Young B, Gales B, Wang HJ, Rastogi A, et~al. (2010) Reduction of
  serum hepcidin by hemodialysis in pediatric and adult patients.
\newblock Clin J Am Soc Nephrol 5: 1010--1014.
\bibAnnoteFile{Zaritsky2010Reduction}

\bibitem{Wang2013Hepatocyte}
Wang CYY, Knutson MD (2013) Hepatocyte divalent metal-ion transporter-1 is
  dispensable for hepatic iron accumulation and non-transferrin-bound iron
  uptake in mice.
\newblock Hepatology : doi:10.1002/hep.26401.
\bibAnnoteFile{Wang2013Hepatocyte}

\bibitem{Kim1995Translational}
Kim HY, Klausner RD, Rouault TA (1995) Translational repressor activity is
  equivalent and is quantitatively predicted by in vitro rna binding for two
  iron-responsive element-binding proteins, irp1 and irp2.
\newblock The Journal of Biological Chemistry 270: 4983--4986.
\bibAnnoteFile{Kim1995Translational}

\bibitem{Ke1998Loops}
Ke Y, Wu J, Leibold EA, Walden WE, Theil EC (1998) Loops and bulge/loops in
  iron-responsive element isoforms influence iron regulatory protein binding.
  fine-tuning of mrna regulation?
\newblock The Journal of Biological Chemistry 273: 23637--23640.
\bibAnnoteFile{Ke1998Loops}

\bibitem{Erlitzki2002Multiple}
Erlitzki R, Long JC, Theil EC (2002) Multiple, conserved iron-responsive
  elements in the 3'-untranslated region of transferrin receptor mrna enhance
  binding of iron regulatory protein 2.
\newblock The Journal of biological chemistry 277: 42579--42587.
\bibAnnoteFile{Erlitzki2002Multiple}

\bibitem{494215}
Kennedy J, Eberhart R (1995) Particle swarm optimization.
\newblock In: Proceedings of the Fourth IEEE International Conference on Neural
  Networks, Perth, Australia. pp. 1942--1948.
\bibAnnoteFile{494215}

\bibitem{condor-hunter}
Litzkow MJ, Livny M, Mutka MW (1988) Condor---a hunter of idle workstations.
\newblock In: 8th International Conference on Distributed Computing Systems.
  pp. 104--111.
\bibAnnoteFile{condor-hunter}

\bibitem{22834945}
Kent E, Hoops S, Mendes P (2012) Condor-copasi: high-throughput computing for
  biochemical networks.
\newblock BMC Systems Biology 6: 91.
\bibAnnoteFile{22834945}

\bibitem{Epsztejn1997Fluorescence}
Epsztejn S, Kakhlon O, Glickstein H, Breuer W, Cabantchik ZI (1997)
  {Fluorescence Analysis of the Labile Iron Pool of Mammalian Cells}.
\newblock Anal Biochem : 31--40.
\bibAnnoteFile{Epsztejn1997Fluorescence}

\bibitem{Haile1989}
Haile DJ, Hentze MW, Rouault TA, Harford JB, Klausner RD (1989) Regulation of
  interaction of the iron-responsive element binding protein with
  iron-responsive {RNA} elements.
\newblock Molecular and cellular biology 9: 5055--5061.
\bibAnnoteFile{Haile1989}

\bibitem{Cozzi2003Role}
Cozzi A (2003) Role of iron and ferritin in tnfa-induced apoptosis in hela
  cells.
\newblock FEBS Lett 537: 187--192.
\bibAnnoteFile{Cozzi2003Role}

\bibitem{Mateo2010Serum}
Mateo I, Infante J, S\'{a}nchez-Juan P, Garc\'{\i}a-Gorostiaga I,
  Rodr\'{\i}guez-Rodr\'{\i}guez E, et~al. (2010) Serum heme oxygenase-1 levels
  are increased in parkinson's disease but not in alzheimer's disease.
\newblock Acta Neurol Scandinavica 121: 136--138.
\bibAnnoteFile{Mateo2010Serum}

\bibitem{Johnson2004Diferric}
Johnson MB, Enns CA (2004) Diferric transferrin regulates transferrin receptor
  2 protein stability.
\newblock Blood 104: 4287--4293.
\bibAnnoteFile{Johnson2004Diferric}

\bibitem{Sassa2004Why}
Sassa S (2004) Why heme needs to be degraded to iron, biliverdin ixalpha, and
  carbon monoxide?
\newblock Antioxidants \& Redox Signaling 6: 819--824.
\bibAnnoteFile{Sassa2004Why}

\bibitem{Pantopoulos1995Differential}
Pantopoulos K, Gray NK, Hentze MW (1995) {Differential regulation of two
  related RNA-binding proteins, iron regulatory protein (IRP) and IRPB.}
\newblock RNA 1: 155--163.
\bibAnnoteFile{Pantopoulos1995Differential}

\bibitem{Wang2003Haemochromatosis}
Wang J, Chen G, Pantopoulos K (2003) The haemochromatosis protein hfe induces
  an apparent iron-deficient phenotype in h1299 cells that is not corrected by
  co-expression of beta 2-microglobulin.
\newblock Biochem J 370: 891--899.
\bibAnnoteFile{Wang2003Haemochromatosis}

\bibitem{Rivera2005Synthetic}
Rivera S, Nemeth E, Gabayan V, Lopez MA, Farshidi D, et~al. (2005) Synthetic
  hepcidin causes rapid dose-dependent hypoferremia and is concentrated in
  ferroportin-containing organs.
\newblock Blood 106: 2196--2199.
\bibAnnoteFile{Rivera2005Synthetic}

\bibitem{Kinobe2006Inhibition}
Kinobe RT, Dercho RA, Vlahakis JZ, Brien JF, Szarek WA, et~al. (2006)
  Inhibition of the enzymatic activity of heme oxygenases by azole-based
  antifungal drugs.
\newblock J Pharmacol Exp Therap 319: 277--284.
\bibAnnoteFile{Kinobe2006Inhibition}

\bibitem{West2000Comparison}
West AP, Bennett MJ, Sellers VM, Andrews NC, Enns CA, et~al. (2000) {Comparison
  of the Interactions of Transferrin Receptor and Transferrin Receptor 2 with
  Transferrin and the Hereditary Hemochromatosis Protein HFE}.
\newblock J Biol Chem 275: 38135--38138.
\bibAnnoteFile{West2000Comparison}

\bibitem{Byrne2010Unique}
Byrne SL, Chasteen ND, Steere AN, Mason AB (2010) The unique kinetics of iron
  release from transferrin: the role of receptor, lobe-lobe interactions, and
  salt at endosomal ph.
\newblock Journal of Molecular Biology 396: 130--140.
\bibAnnoteFile{Byrne2010Unique}

\bibitem{Pimstone1971Inducible}
Pimstone NR, Engel P, Tenhunen R, Seitz PT, Marver HS, et~al. (1971) Inducible
  heme oxygenase in the kidney: a model for the homeostatic control of
  hemoglobin catabolism.
\newblock J Clin Investigation 50: 2042--2050.
\bibAnnoteFile{Pimstone1971Inducible}

\bibitem{Bao2010Plasma}
Bao W, Song F, Li X, Rong S, Yang W, et~al. (2010) Plasma heme oxygenase-1
  concentration is elevated in individuals with type 2 diabetes mellitus.
\newblock PLoS One 5: e12371.
\bibAnnoteFile{Bao2010Plasma}

\bibitem{Shayeghi2005Identification}
Shayeghi M, Latunde-Dada GO, Oakhill JS, Laftah AH, Takeuchi K, et~al. (2005)
  {Identification of an intestinal heme transporter.}
\newblock Cell 122: 789--801.
\bibAnnoteFile{Shayeghi2005Identification}

\bibitem{Tamura2006Functional}
Tamura A, Watanabe M, Saito H, Nakagawa H, Kamachi T, et~al. (2006) Functional
  validation of the genetic polymorphisms of human atp-binding cassette (abc)
  transporter abcg2: identification of alleles that are defective in porphyrin
  transport.
\newblock Mol Pharmacol 70: 287--296.
\bibAnnoteFile{Tamura2006Functional}

\bibitem{Cairo1998Lack}
Cairo G, Tacchini L, Pietrangelo A (1998) Lack of coordinate control of
  ferritin and transferrin receptor expression during rat liver regeneration.
\newblock Hepatol 28: 173--178.
\bibAnnoteFile{Cairo1998Lack}

\bibitem{Summers1974}
Summers M, Worwood M, Jacobs A (1974) Ferritin in normal erythrocytes,
  lymphocytes, polymorphs, and monocytes.
\newblock British Journal of Haematology 28: 19--26.
\bibAnnoteFile{Summers1974}

\bibitem{SalterCid1999Transferrin}
Salter-Cid L, Brunmark A, Li Y, Leturcq D, Peterson PA, et~al. (1999)
  Transferrin receptor is negatively modulated by the hemochromatosis protein
  hfe: implications for cellular iron homeostasis.
\newblock Proc Natl Acad Sci USA 96: 5434--5439.
\bibAnnoteFile{SalterCid1999Transferrin}

\bibitem{Sibille1988Interactions}
Sibille JC, Kondo H, Aisen P (1988) Interactions between isolated hepatocytes
  and kupffer cells in iron metabolism: a possible role for ferritin as an iron
  carrier protein.
\newblock Hepatol 8: 296--301.
\bibAnnoteFile{Sibille1988Interactions}

\bibitem{Swinkels2008Advances}
Swinkels DW, Girelli D, Laarakkers C, Kroot J, Campostrini N, et~al. (2008)
  Advances in quantitative hepcidin measurements by time-of-flight mass
  spectrometry.
\newblock PloS One 3: 7.
\bibAnnoteFile{Swinkels2008Advances}

\bibitem{Riedel1999HFE}
Riedel HD, Muckenthaler MU, Gehrke SG, Mohr I, Brennan K, et~al. (1999) Hfe
  downregulates iron uptake from transferrin and induces iron-regulatory
  protein activity in stably transfected cells.
\newblock Blood 94: 3915--3921.
\bibAnnoteFile{Riedel1999HFE}

\bibitem{vanDijk2008Serum}
van Dijk BA, Laarakkers CM, Klaver SM, Jacobs EM, van Tits LJ, et~al. (2008)
  Serum hepcidin levels are innately low in hfe-related haemochromatosis but
  differ between c282y-homozygotes with elevated and normal ferritin levels.
\newblock British J Haematol 142: 979--985.
\bibAnnoteFile{vanDijk2008Serum}

\end{thebibliography}

\clearpage

\section*{Figure Legends}

\clearpage

\section*{Tables}

%%Table of initial concentrations of metabolites and sources
\begin{table}[!ht]
\caption{ {\bf Initial Conditions.} Initial concentrations of all metabolites and the source for their value.}
\begin{tabular}{|c|c|c|}
\hline
                        Parameter & Initial Concentration (mol/l) & Source\\
			\hline
			LIP & 1.3E-6 & \cite{Epsztejn1997Fluorescence}\\\hline
			FPN1 & 1E-9 & \\\hline
			IRP & 1.16E-06 & \cite{Haile1989}\\\hline
			HAMP & 5E-9 & \cite{Zaritsky2010Reduction}\\\hline
			haeme & 1E-9 & \\\hline
			2(Tf-Fe)-TfR1\_Internal & 0 & \\\hline
			2(Tf-Fe)-TfR2\_Internal & 0 & \\\hline
			Tf-Fe-TfR2\_Internal & 0 & \\\hline
			Tf-Fe-TfR1\_Internal & 0 & \\\hline
			Tf-TfR1\_Internal & 0 & \\\hline
			Tf-TfR2\_Internal & 0 & \\\hline
			Fe-FT & 0 & \\\hline
			FT & 1.66E-10 & \cite{Cozzi2003Role}\\\hline
			HO-1 & 3.56E-11 & \cite{Mateo2010Serum}\\\hline
			FT1 & 0 & \\\hline
			Tf-Fe\_intercell & 5E-6 & fixed, \cite{Johnson2004Diferric}\\\hline
			TfR & 4E-7 & \cite{Chloupkova2010Stoichiometries}\\\hline
			Tf-Fe-TfR1 & 0 & \\\hline
			HFE & 2e-07 & \cite{Chloupkova2010Stoichiometries}\\\hline
			HFE-TfR & 0 & \\\hline
			HFE-TfR2 & 0 & \\\hline
			Tf-Fe-TfR2 & 0 & \\\hline
			2(Tf-Fe)-TfR1 & 0 & \\\hline
			2HFE-TfR & 0 & \\\hline
			2HFE-TfR2 & 0 & \\\hline
			2(Tf-Fe)-TfR2 & 0 & \\\hline
			TfR2 & 3E-6 & \cite{Chloupkova2010Stoichiometries}\\\hline
			haeme\_intercell & 1e-07 & \cite{Sassa2004Why}\\\hline
\end{tabular}
\begin{flushleft}
\end{flushleft}
\label{tab:tableOfParams}
 \end{table}

%%Table of reaction parameters and source
\clearpage
\begin{small}
\begin{longtable}{|p{3cm}|p{5cm}|p{3cm}|p{4cm}|c|}
\caption{Reactions, rate laws and kinetic constant values.}
\\
\hline
Name & Reaction & Rate law & Parameters & Source\\
\hline
\endfirsthead
\multicolumn{5}{c}%
{\tablename\ \thetable\ -- \textit{Continued from previous page}} \\
\hline
Name & Reaction & Rate Law & Parameters & Source\\
\hline
\endhead
\hline \multicolumn{5}{r}{\textit{Continued on next page}} \\
\endfoot
\hline
				Fpn Export & LIP $\to$ Tf-Fe\_intercell;  FPN1 & Hill Function $\to$ & a=15 mol s$^{-1}$, $n_H=1$, k=1E-6 mol & \cite{Sarkar2003}\\
				TfR1 expression &  $\to$ TfR;  IRP & Hill Function $\to$ & a=6e-12 s$^{-1}$, $n_H=1$, k=1e-6 mol & \cite{Chloupkova2010Stoichiometries} \\
				TfR1 degradation & TfR $\to$  & Mass action  & k=8.37e-06 s$^{-1}$ & \cite{Johnson2004Diferric} \\
				Ferroportin Expression &  $\to$ FPN1;  IRP & Hill Function -$|$ & a=4e-9 s$^{-1}$, $n_H=1$, $K=$1e-6 mol &\\
				IRP expression &  $\to$ IRP;  LIP & Hill Function -$|$ & a=4e-11 s$^{-1}$, $n_H=1$, K=1e-6 mol & \cite{Pantopoulos1995Differential} \\
				IRP degradation & IRP $\to$  & Mass action & k = 1.59e-5  s$^{-1}$& \cite{Pantopoulos1995Differential} \\
				Fpn degradation Hepc& FPN1 $\to$ ;  HAMP & Hill Function $\to$  & a=2.315E-5 s$^{-1}$, $n_H=1$, K=1e-9 mol &\\
				HFE degradation & HFE $\to$  & Mass action & k = 6.418e-5  s$^{-2}$& \cite{Wang2003Haemochromatosis} \\
				HFE expression &  $\to$ HFE & Constant flux & v = 2.3469e-11 mol (l $\cdot$ s)$^{-1}$ & \cite{Wang2003Haemochromatosis} \\
				TfR2 expression &  $\to$ TfR2 & Constant flux & v = 2e-11 mol (l $\cdot$ s)$^{-1}$& \cite{Chloupkova2010Stoichiometries} \\
				TfR2 degradation & TfR2 $\to$ ;  Tf-Fe\_intercell & Hill Function -$|$ & a=3.2e-05 s$^{-1}$, $n_H=1$, K=2.5e9 mol & \cite{Chloupkova2010Stoichiometries} \\
				Hepcidin expression &  $\to$ HAMP;  2HFE-TfR2 2(Tf-Fe)-TfR2& Hill Function $\to$ & a=5e-12 s$^{-1}$, $n_H=5$, K=1.35e-07 mol, a=5e-12 mol  s$^{-1}$, K=6e-7 mol & \cite{Zaritsky2010Reduction} \\
				Hepcidin degradation & HAMP $\to$  & Mass action & k = 9.63e-05 s$^{-1}$ & \cite{Rivera2005Synthetic} \\
				Heme oxygenation & Heme $\to$ LIP;  HO-1 & Henri-Michaelis-Menten & kcat=17777.7  s$^{-1}$, Km=2e-6 mol l$^{-1}$ & \cite{Kinobe2006Inhibition} \\
				HFE TfR1 binding & HFE + TfR $\to$ HFE-TfR & Mass action & k = 1.102e+06 l (mol $\cdot$ s)$^{-1}$ & \cite{West2000Comparison} \\
				HFE TfR1 release & HFE-TfR $\to$ HFE + TfR & Mass action & k = 0.08 s$^{-1}$ & \cite{West2000Comparison} \\
				TfR1 binding & Tf-Fe\_intercell + TfR $\to$ Tf-Fe-TfR1 & Mass action & k = 837400 l (mol $\cdot$ s)$^{-1}$& \cite{West2000Comparison} \\
				TfR1 release & Tf-Fe-TfR1 $\to$ Tf-Fe\_intercell + TfR & Mass action & k = 9.142e-4 s$^{-1}$  & \cite{West2000Comparison} \\
				HFE TfR2 binding & $2 * $HFE + TfR2 $\to$ 2HFE-TfR2 & Mass action & k = 3.9438e+11 l$^{2}$ (mol$^{2}$ $\cdot$ s)$^{-1}$ & \\
				HFE TfR2 release & 2HFE-TfR2 $\to$ 2 * HFE + TfR2 & Mass action & k = 0.0018  s$^{-1}$ &  \\
				TfR2 binding & Tf-Fe\_intercell + TfR2 $\to$ Tf-Fe-TfR2 & Mass action & k = 222390 l (mol $\cdot$ s)$^{-1}$& \cite{West2000Comparison} \\
				TfR2 release & Tf-Fe-TfR2 $\to$ Tf-Fe\_intercell + TfR2 & Mass action & k = 0.0061 s$^{-1}$ & \cite{West2000Comparison} \\
				TfR1 binding 2 & Tf-Fe-TfR1 + Tf-Fe\_intercell $\to$ 2(Tf-Fe)-TfR1 & Mass action & k = 121400 l (mol $\cdot$ s)$^{-1}$& \cite{West2000Comparison} \\
				TfR1 release 2 & 2(Tf-Fe)-TfR1 $\to$ Tf-Fe-TfR1 + Tf-Fe\_intercell & Mass action & k = 0.003535 s$^{-1}$ & \cite{West2000Comparison} \\
				HFE TfR1 binding 2 & HFE-TfR + HFE $\to$ 2HFE-TfR & Mass action & k = 1.102e+06 l (mol $\cdot$ s)$^{-1}$& \cite{West2000Comparison} \\
				HFE TfR1 release 2 & 2HFE-TfR $\to$ HFE-TfR + HFE & Mass action & k = 0.08 s$^{-1}$ & \cite{West2000Comparison} \\
				TfR2 binding 2 & Tf-Fe-TfR2 + Tf-Fe\_intercell $\to$ 2(Tf-Fe)-TfR2 & Mass action & k= 69600 l (mol $\cdot$ s)$^{-1}$& \cite{West2000Comparison} \\
				TfR2 release 2 & 2(Tf-Fe)-TfR2 $\to$ Tf-Fe-TfR2 + Tf-Fe\_intercell & Mass action & k = 0.024 s$^{-1}$ & \cite{West2000Comparison} \\
				TfR1 iron internalisation & 2(Tf-Fe)-TfR1 $\to$ 4(LIP) + TfR & Mass action & k = 0.8333 l$\cdot$s$^{-1}$  & \cite{Byrne2010Unique} \\
				TfR2 iron internalisation & 2(Tf-Fe)-TfR2 $\to$ 4(LIP)-TfR2 & Mass action & k = 0.8333 l$\cdot$s$^{-1}$ & \cite{Byrne2010Unique} \\
				outFlow & LIP $\to$  & Mass action (irreversible) & k = 4e-04 s$^{-1}$ & \\
				Ferritin Iron binding & LIP + FT $\to$ Fe-FT & Mass action & k = 4.71e+10 l (mol $\cdot$ s)$^{-1}$& \cite{Salgado2010Mathematical} \\
				Ferritin Iron release & Fe-FT $\to$ LIP + FT & Mass action & k = 22922 s$^{-1}$  & \cite{Salgado2010Mathematical} \\
				Ferritin Iron internalisation & Fe-FT $\to$ FT1 + FT & Mass action & k = 108000 s$^{-1}$ & \cite{Salgado2010Mathematical} \\
				Ferritin internalised iron release & FT1 $\to$ LIP;  FT1 FT & Kloss Hill & kloss = 13.112 s$^{-1}$  & \cite{Salgado2010Mathematical} \\
				ferritin expression &  $\to$ FT;  IRP & Hill Function -$|$ & a=2.312e-13 s$^{-1}$, $n_H=1$, K=1e-06 mol & \cite{Cozzi2003Role} \\
				HO1 Degradation & HO-1 $\to$  & Mass action  & k= 3.209e-05 s$^{-1}$ & \cite{Pimstone1971Inducible} \\
				HO1 Expression &  $\to$ HO-1;  Heme & Hill Function $\to$ & a=2.1432e-15 s$^{-1}$, K=1e-9 mol& \cite{Bao2010Plasma} \\
				Ferritin Degradation Full & FT $\to$  & Mass action & k = 1.203e-05 s$^{-1}$& \cite{Salgado2010Mathematical} \\
				Heme uptake & Heme\_intercell $\to$ Heme & Henri-Michaelis-Menten & Km=1.25e-4 mol, v=1.034e-5 mol s$^{-1}$ & \cite{Shayeghi2005Identification} \\
				Heme export & Heme $\to$ Heme\_intercell & Henri-Michaelis-Menten & Km=1.78e-05 mol, v=2.18e-5 mol s$^{-1}$& \cite{Tamura2006Functional} \\
				Ferritin Degradation Full Iron Release & FT1 $\to$ LIP;  FT1 FT & Mass Action Ferritin & k=1.203e-05 s$^{-1}$ & \cite{Salgado2010Mathematical} \\
				HFE-TfR degradation & 2HFE-TfR $\to$  & Mass action & k=8.37e-07 s$^{-1}$&  \\
				HFE-TfR2 degradation & 2HFE-TfR2 $\to$  & Mass action  & k=8.37e-07 s$^{-1}$&  \\
\label{tab:reaction_parameters}
\end{longtable}
\end{small}

\begin{table}[!ht]
\caption{
{\bf Steady State Validation} --- Comparison between model and experimental observations. IRP, Ferritin and TfR are expressed in particles per cell assuming a cellular volume of $10^{-12}$ l. Iron per Ferritin is a ratio. }
\begin{tabular}{|c|c|c|c|}
\hline
			Metabolite & Model & Experimental & Reference \\
			\hline
			Labile Iron Pool & $0.804\mu M$ & $0.2-1.5 \mu M$ & \cite{Epsztejn1997Fluorescence}\\
			Iron Response Protein & $836000$ cell$^{-1}$ & $\sim700000$ cell$^{-1}$ & \cite{Cairo1998Lack}\\
			Ferritin & $4845$ cell$^{-1}$ & $3000-6000$ cell$^{-1}$ (mRNA), $2.5-54600$ cell$^{-1}$(protein) & \cite{Cairo1998Lack}, \cite{Summers1974}\\
			TfR & $1.74\times10^{5}$ cell$^{-1}$ & $1.6-2\times10^{5}$ cell$^{-1}$ & \cite{SalterCid1999Transferrin}\\
			TfR2 & $4.63\times\left[\textrm{TfR1}\right]$ & $4.5-6.1\times\left[\textrm{TfR1}\right]$ & \cite{Chloupkova2010Stoichiometries}\\
			Iron per Ferritin & 2272 average & ~2400 &\cite{Sibille1988Interactions}\\
			Hepcidin & $5.32$nM & $3.5-8.3$nM  & \cite{Swinkels2008Advances}\\
			\hline
			\hline
			Reaction & Model & Experimental & Reference\\
			\hline
			TBI import rate& $2.67\mu M$ s$^{-1}$ & $2.08\mu M$ s$^{-1}$ & \cite{Chua2010Iron}\\
\hline
\end{tabular}
\label{tab:tableOfSS}
 \end{table}

\begin{table}[!ht]
\caption{ {\bf HFE Knockout Validation} --- The simulation of type-1 hereditary haemochromatosis closely matches experimental findings at steady state.}
\begin{tabular}{|l|c|c|r|}
		  \hline
			Metabolite & Model & Experiment & Reference \\
			\hline
			IRP & - & - & \cite{Riedel1999HFE}\\
			LIP & + & + & \cite{Riedel1999HFE}\\
			HAMP & - & - & \cite{vanDijk2008Serum}\\
			TfR2 & + & + & \cite{Robb2004Regulation}\\
			\hline
			Reaction & Model & Experimental & Reference\\
			\hline
			TfR1/2 iron import& + & + &  \cite{Riedel1999HFE}\\
			FT expression& + & + & \cite{Riedel1999HFE}\\
			TfR expression& - & - & \cite{Riedel1999HFE}\\
			FPN expression& $\approx$ & = & \cite{Ludwiczek2005Regulatory}\\
                        \hline
\end{tabular}
\label{table:tableHFEKO}
 \end{table}

\begin{table}[!ht]
\caption{ {\bf Metabolic Control Analysis} --- Concentration control coefficients  for the labile iron pool.}
\begin{tabular}{|c|l|l|l|}
		 	\hline
			\multirow{2}{*}{ Reaction} & \multirow{2}{*}{Local} & \multicolumn{2}{|c|} {Global} \\
			 & & Minimum & Maximum\\
			\hline
                        TfR2 expression&0.894573&0.515971&1.41255\\
                        Fpn Export&-0.825483&-0.924&-0.698754\\
                        TfR2 binding&0.569815&0.298433&0.901285\\
                        TfR2 degradation&-0.563132&-0.898362&-0.293111\\
                        Fpn degradation&0.351397&0.186176&0.50289\\
                        Ferroportin Expression&-0.351397&-0.502317&-0.176245\\
                        HFE expression&-0.313525&-0.623067&0.346532\\
                        TfR1 expression&0.259758&0.0652&0.496352\\
                        TfR1 binding&0.259436&0.06577&0.497636\\
                        TfR1 degradation&-0.258004&-0.503067&-0.0657364\\
                        IRP expresion&0.209893&0.0748546&0.300039\\
                        IRP degradation&-0.209893&-0.347477&-0.0753367\\
                        HFE-TfR2 degradation&-0.0341692&-0.684936&0.000229851\\
                        Hepcidin expression&0.0283652&0.0004375&0.6553120\\
                        Hepcidin degradation&-0.0283652&-0.791216&-0.000576136\\
                        HFE degradation&0.0162284&-0.0259426&0.0386967\\
                        TfR2 binding 2&0.0100938&0.298433&0.901285\\
                        TfR2 release&-0.0100938&-0.0194113&-0.00434313\\
                        HFE TfR2 binding&-0.00668253&-0.0187053&0.0218869\\
                        HFE TfR2 release&0.0063856&-0.0205303&0.018034\\
                        TfR2 iron internalisation&-0.00335169&-0.156882&0.000557494\\
                        HFE TfR1 binding&-0.00143167&-0.0120993&0.0000742\\
                        HFE TfR1 release&0.00143166&0.0000760&0.0121124\\
                        HFE TfR1 binding 2&-0.00143166&-0.0121238&-0.0000739\\
                        HFE TfR1 release 2&0.00143165&0.0000738&0.0121135\\
                        HFE-TfR degradation&-0.00143165&-0.0121249&-0.0000737\\
\hline
\end{tabular}
\label{table:MCA}
 \end{table}

\begin{table}[!ht]
\caption{ {\bf Metabolic Control Analysis} --- Concentration control coefficients for hepcidin.}
\begin{tabular}{|c|l|l|l|}
		 	\hline
			\multirow{2}{*}{ Reaction} & \multirow{2}{*}{Local} & \multicolumn{2}{|c|} {Global} \\
			 & & Minimum & Maximum\\
			\hline
                        Hepcidin expression&1.00002&0.512257&1.487664\\
                        Hepcidin degradation&-1.00002&-1.00027&-0.999001\\
                        HFE-TfR2 degradation&-0.956041&-1.3943&-0.380497\\
                        HFE expression&0.9131&0.274035&1.30051\\
                        TfR2 expression&0.243052&0.0984356&0.486305\\
                        TfR2 degradation&-0.153001&-0.293528&-0.0638787\\
                        TfR2 binding&0.128436&0.0558287&0.273304\\
                        TfR2 iron internalisation&-0.128062&-0.272967&-0.0557919\\
                        HFE degradation&-0.047263&-0.102578&-0.0122656\\
                        HFE TfR2 binding&0.0245645&0.00630724&0.0573883\\
                        HFE TfR2 release&-0.023473&-0.0557905&-0.00602681\\
                        TfR2 binding 2&0.00227514&0.000811688&0.00589495\\
                        TfR2 release&-0.00227514&-0.00589437&-0.000812498\\
                        HFE TfR1 binding&-0.00093303&-0.00728765&-5.22895e-05\\
                        HFE TfR1 release&0.000933028&4.84169e-05&0.00697082\\
                        HFE TfR1 binding 2&-0.000933028&-0.0073373&-5.31758e-05\\
                        HFE TfR1 release 2&0.000933018&5.3417e-05&0.00731269\\
                        HFE-TfR degradation&-0.000933018&-0.00733725&-5.69006e-05\\
                        TfR1 expression&-0.000796332&-0.00607511&-4.36181e-05\\
                        TfR1 degradation&0.000790955&4.53395e-05&0.00623214\\
                        IRP expresion&-0.000544238&-0.00281211&-4.71681e-05\\
                        IRP degradation&0.000544238&4.16666e-05&0.00351147\\
                        Fpn export&-0.00045206&-0.00277642&-4.33404e-05\\
                        Fpn degradation&0.000192436&1.47897e-05&0.00153538\\
                        Ferroportin expression&-0.000192436&-0.00153463&-1.41905e-05\\
                        TfR1 binding&0.000142075&3.78713e-06&0.00137383\\
                        TfR2 release 2&-6.36921e-05&-0.000176906&-2.18216e-05\\
\hline
\end{tabular}
\label{table:MCAHepc}
 \end{table}

\begin{table}[!ht]
\caption{ {\bf Metabolic Control Analysis} --- Flux-control coefficients for the iron export out of the liver compartment.}
\begin{tabular}{|c|l|l|l|}
\hline
			\multirow{2}{*}{ Reaction} & \multirow{2}{*}{Local} & \multicolumn{2}{|c|} {Global} \\
			 & & Minimum & Maximum\\
\hline
TfR2 expression&0.910944&0.449405&1.38521\\
TfR2 binding&0.581149&0.285737&0.867434\\
TfR2 degradation&-0.573438&-0.858215&-0.278218\\
HFE expression&-0.353566&-0.669513&-0.187987\\
TfR1 expression&0.266964&0.0676606&0.510467\\
TfR1 binding&0.266601&0.0675083&0.51963\\
TfR1 degradation&-0.265162&-0.51689&-0.0669265\\
IRP expresion&0.182446&0.063823&0.310888\\
IRP degradation&-0.182446&-0.313848&-0.0656558\\
Fpn Export&0.151719&0.0626056&0.271594\\
Ferroportin Expression&0.0645849&0.0189112&0.149717\\
Fpn degradation&-0.0645849&-0.149993&-0.0189094\\
HFE degradation&0.0183009&0.00812358&0.0401559\\
TfR2 release&-0.0102946&-0.018781&-0.00414945\\
TfR2 binding 2&0.0102946&0.00414543&0.0187846\\
HFE TfR2 binding&-0.0077113&-0.0191638&0.00292511\\
HFE TfR2 release&0.00736866&-0.00282598&0.0186586\\
Hepcidin expression&-0.00521336&-0.1785377&-0.0000387334\\
Hepcidin degradation&0.00521336&5.77312e-05&0.224586\\
HFE-TfR2 degradation&-0.00226218&-0.0183295&0.19571\\
HFE TfR1 binding&-0.00143917&-0.0119501&-7.50839e-05\\
HFE TfR1 release&0.00143917&7.49065e-05&0.0119095\\
HFE TfR1 binding 2&-0.00143917&-0.0114124&-7.49686e-05\\
HFE TfR1 release 2&0.00143915&7.49046e-05&0.0116242\\
\hline
\end{tabular}
\label{table:MCAFlux}
\end{table}

\end{document}